# Light Absorption Coefficients of Ionic Liquids under Electric Field*

Ji Zhou (周吉)[1], Shikui Dong (董士奎)[2], Zhihong He (贺志宏)[2], Julius Caesar Puoza[3], and Yanhu Zhang (张彦虎)[4†]

[1] *Beijing Institute of Space Mechanics & Electricity, China Academy of Space Technology, Beijing 100094, P. R. China*

[2] *School of Energy Science and Engineering, Harbin Institute of Technology, Harbin 150001, P. R. China*

[3] *Department of Mechanical Engineering, Sunyani Technical University, P. O. Box 206, Sunyani, Ghana*

[4] *Adv. Manufacturing & Equipment Institute, Jiangsu University, Zhenjiang 212013, P. R. China*

Ionic liquids have attracted a lot of research attention for their applications in novel optoelectronic structures and devices as an optical means of regulating electricity. Although the electro-optic effect of ionic liquids is mentioned in some literatures, a quantitative testing and analysis is hardly found in light absorption coefficients of ionic liquids under electric field. In the present study, an experimental apparatus was designed to measure the absorption coefficients of ionic liquids under different electric fields. Five groups of imidazole ionic liquids were experimentally investigated and an inversion was performed to determine the spectral absorption coefficients of the imidazole ionic liquids under electric fields. Different intensities with multiple interface refractions and reflections were also considered, and the various measurement errors were analyzed through uncertainties propagation analysis. Spectral absorption of ionic liquids from 300 nm to 2500 nm was obtained and the absorption coefficients were retrieved. It was found that the absorption behavior of ionic liquids changed in some frequency bands under electric field. The experimental results showed that the absorption coefficient of the ionic liquid increases with the voltage at 1520 nm and 1920 nm. The change rate was affected by the types of anions and cations in the ionic liquid and the diffusion rate of the ions therein. This study provides illustrations for the ionic liquid-based electro-optical regulation in terms of physical property parameters and the testing technique.

**Keywords:** Ionic liquid; absorption coefficient; electro-optical property; uncertainty propagation analysis

**PACS:** 78.40.-q, 78.20.Ci, 33.57.+c, 78.20.Jq, 78.30.cd

## 1. Introduction

Ionic liquid, called room temperature molten salt, is a new kind of electro-optical material which differ from dielectrics and semiconductors. It has excellent electric, magnetic, acoustic, thermal, and optical properties. Ionic liquid can be mixed or combined with some hard materials to form various novel superconducting materials that having unique electro-optical, magneto-optical, and/or thermo-optical properties.

* Project supported by National Natural Science Foundation of China (Grant No.51576054, 51705210)
† Corresponding author. E-mail: zhyh@ujs.edu.cn

Ionic liquid has been designated as a promising functional material in the 21st century. Relevant studies [1,2] showed that an ionic liquid enables wide-spectrum electro-optical regulation by optical means. The optical transmission behavior of the ionic liquid in an applied field is of great significance in promoting the application in the electro-optical regulation field.

Room-temperature imidazole ionic liquid has received considerable attention in recent years because it can be substituted for volatile organic compounds and is widely used in synthesis, catalysis, electrochemistry, and optical physics [3-6]. The optical absorption properties have been studied by many researchers. For instance, Du et al. [7] explored the explicit correlation between the structural and optical properties of an imidazolium amino acid-based ionic liquid. Their results showed that the absorption behaviours of imidazolium-based ionic liquids are sensitive to the local heterogeneous environments. Song et al. [8] measured the UV spectrum of 1-methyl-3-butyl imidazole tetrafluoroborate in water within the spectral range of 200-400nm and found that the maximum light absorption wavelength was determined to be 221nm. Yang [9] and Yu et al. [10] also explored the UV-spectrum absorption behaviour of 1-methyl-3-butyl imidazole nitrate in ethanol and water. Paul et al. [11-13] presented the light absorption and fluorescence behaviours of a series of imidazole ionic liquids within the ultraviolet and visible spectra. Their findings indicated that the light absorption of PF6/BF4-based imidazole ionic liquids is indeed not negligible, differing from conventional transparent media. Furthermore, the maximum fluorescence values depended heavily on the wavelength of the excitement wave. Jo et al. [14] studied the emissivity differences of ionic liquids of pyridine (quaternary ammonium) salts with different radicals at position 2 under different wavelengths. Zhang et al. [15] experimentally examined the radiation properties of [Hmim][Tf2N] ionic liquid and its nanofluid. Their findings demonstrated that [Hmim][Tf2N] was nearly transparent within the visible spectrum, and its light absorption property was dramatically improved due to some nanoparticles scattered within. Ansari et al. [16] explored the light absorption and emission properties of binary lanthanide/nitrate complex and ternary lanthanide/nitrate/chloride complex in a water-bearing ionic liquid.

Interesting, Hu et al. [17] developed a composite material with unique electro-optical properties. Under an electric field, the material can provide a chiral radical density gradient distribution, resulting in broadband reflection and three states of transition which are transparency, diffuse reflection, and mirror reflection. This technology has a potential application for smart glass and electronic paper. Besides, Wang et al. [2] revealed that ultra-high-density electron holes and their electro-optical regulation within the far infrared and visible spectra, which provides an approach to develop the large-area electrochromic smart window used in energy-saving buildings and vehicles. Nakano et al. [18] invented a $VO_2$-based infrared-sensitive field-effect device, which can serve as an electric switch for light transmittance and conventional current for the voltage regulation and thermal cutting filter. Reddy et al. [19] developed a novel electrochromic device that was synthesized in a hydrophobic ionic liquid and each cathode and anode consisted of a thermally stable plasma substrate. He et al. [20] studied the electro-optic effect of the imidazole ionic liquid in the optofluidic waveguide. Results showed that the absorption behavior was attributed to the imidazole part and relevant structures. In addition, both the regulation range and response speed increase with the voltage as well as the electrical conductivity of the ionic liquid.

From the literature of the above-mentioned research works, very little is known on the accuracy and error of absorption coefficient measurements in the scope of light absorption behavior of ionic liquids. Specifically, there is no report on the analysis of propagated uncertainties in measurement results affected by multiple measurement factors. Moreover, knowledge of the electro-optic property changes of ionic liquids in an applied electric field is limited. For their optical parameters in an electrical field, it is currently impossible to quantitatively identify the impact of the types of cations and anions and carbon chain length on the optical properties of ionic liquids on their absorption coefficients. This prevents further development and application of ionic liquids.

In this study, an inversion was performed on the spectral transmittance measurements to identify the absorption coefficient versus external voltage relation, with multiple interface refractions and reflections considered. Additionally, an analysis of propagated uncertainties was carried out to identify the effects of different errors on the light absorption coefficient inversion results. Lastly, an experimental measurement program based on the transmission method was designed for $C_3MImI$ (methyl propyl imidazole iodide) and its four derivatives including $C_5MImI$ (methyl pentyl imidazole iodide), $C_4MImI$ (methyl butyl imidazole iodide), $C_3MImBr$ (methyl propyl imidazole bromide), and $C_3MImBF_4$ (methyl propyl imidazole tetrafluoroborate). This study will have great significance for accurate measurements of ionic liquids' absorption coefficients and promoting their application in the electro-optic regulation field.

## 2. Theoretical Analysis
### 2.1. Ionic Liquid's Absorption Coefficient Inversion

In the experiment of spectral transmittance, an inversion model was necessary to determine the absorption coefficient. During the process, the impacts of the sample tank and interface refractions needed to be considered. A physical model was built as shown in Fig. 1. The light transmission process of the empty glass tank and that of the liquid-containing glass tank were illustrated by single-layer model and double-layer model, respectively.

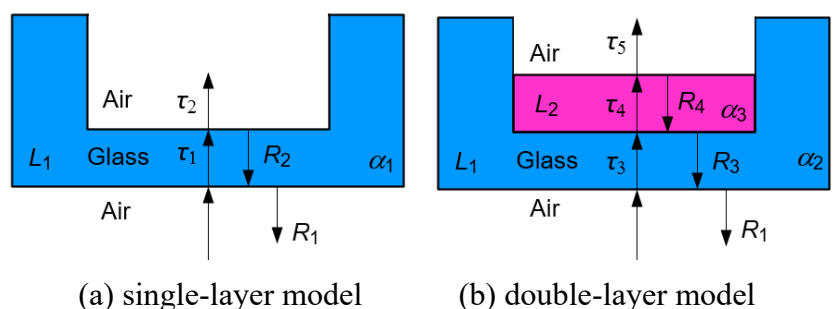

(a) single-layer model    (b) double-layer model

Fig. 1 Illustration of physical models for transmittance.

Note: $\tau_1$ – transmittance of lower surface of glass tank; $\tau_2$ – transmittance of upper surface of glass tank; $\tau_3$ – transmittance of lower surface of liquid-containing glass tank; $\tau_4$ – transmittance

of upper surface of liquid-containing glass tank; $\tau_5$ – transmittance of upper surface of liquid-containing glass tank; $R_1$ – reflectance between glass and air when light enters from air into glass; $R_2$ – reflectance between air and glass when light enters from glass into air; $R_3$ – reflectance between glass and liquid; $R_4$ – reflectance between air and liquid.

Light always struck perpendicularly on the interfaces during the entire measurement process. Then, the reflectance of light into the glass from the air is equal to that of light into the air from the glass. For the double-layer model, the reflectance can be expressed by formula (1).

$$\begin{cases} R_1 = R_2 = \left[(n_{glass} - n_{air})/(n_{glass} + n_{air})\right]^2 \\ R_3 = \left[(n_{glass} - n_{liquid})/(n_{glass} + n_{liquid})\right]^2 \\ R_4 = \left[(n_{liquid} - n_{air})/(n_{liquid} + n_{air})\right]^2 \end{cases} \quad (1)$$

where $n_{liquid}$ is the refractive index of the measured liquid, $n_{glass}$ is the refractive index of the sample tank, and $n_{air}$ is the refractive index of the air and is assumed to be 1 in this research. During the absorption coefficient inversion process, multiple reflections and refractions in the media are considered.

The transmittance of the empty glass tank (without liquid) is calculated by [21]

$$\tau_2 = (1-\alpha_1)(1-R_1)(1-R_2)/\left[1 - R_1 R_2 (1-\alpha_1)^2\right] \quad (2)$$

When the glass tank contains liquid, multiple interfaces are involved, and the physical model is shown in Fig. 2. The model consists of two layers ($b$, $b+1$), and the following can be determined and obtained through derivation operation.

$$\begin{cases} F_{A_b}^{B_b} = \exp(-k_b L_b)/\left[1 - R_B R_A \exp(-2k_b L_b)\right] \\ F_{B_b}^{B_b} = R_A \exp(-2k_b L_b)/\left[1 - R_B R_A \exp(-2k_b L_b)\right] \\ F_{B_{b+1}}^{B_{b+1}} = R_C \exp(-2k_{b+1} L_{b+1})/\left[1 - R_B R_C \exp(-2k_{b+1} L_{b+1})\right] \\ F_{B_{b+1}}^{C_{b+1}} = \exp(-k_{b+1} L_{b+1})/\left[1 - R_B R_C \exp(-2k_{b+1} L_{b+1})\right] \end{cases} \quad (3)$$

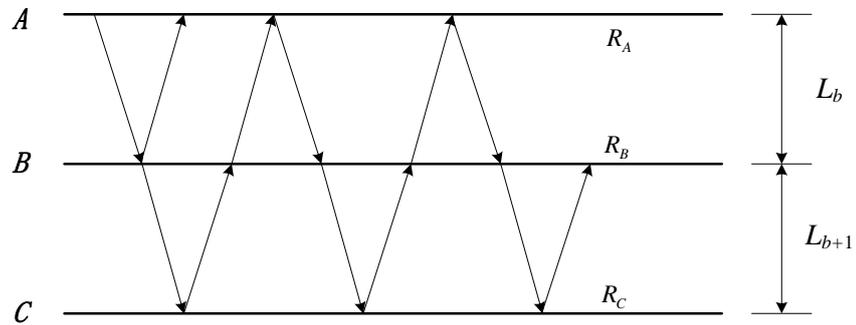

Fig. 2 Double-layer radiant intensity transfer model

Note: $A$ is the interface between the air and the glass, $B$ is the interface between the liquid and the glass, and $C$ is the interface between the liquid and the air. The subscript $b$ means the #$b$ medium layer, $b+1$ means the #$b+1$ medium layer. $F_{A_b}^{B_b}$ means the proportion of the energy

that passes through the interface $A$ and arrives at the upper surface of the interface $B$. The meaning of other parameters is similar.

The total proportion of energy that departs from interface $A_b$ and arrives at interface $C_{b+1}$ can be obtained as follows [22]:

$$E_{A_b,b\sim b+1}^{C_{b+1}} = \tau_{b,b+1} F_{A_b}^{B_b} F_{B_{b+1}}^{C_{c+1}} / \left(1 - \beta_2^S\right) \qquad (4)$$

where $\beta_2^S = \tau_{b,b+1} F_{B_{b+1}}^{B_{b+1}} \tau_{b+1,b} F_{B_b}^{B_b}$, $\tau_{b,b+1}$ means the transmittance when the light enters the medium $b+1$ from the medium $b$ and $\tau_{b+1,b}$ means the transmittance when the light enters the medium $b$ from the medium $b+1$. $R_A$, $R_B$, and $R_C$ are respectively the reflectance on the three interfaces A, B, and C. $R_A=R_1=R_2$, $R_B=R_3$, and $R_C=R_4$. In addition, $\tau_{b,b+1}=\tau_{b+1,b}=1-R_B$. When the glass tank contains liquid, the overall transmittance measured by the spectrograph can be expressed by

$$\tau_5 = E_{A_b,b\sim b+1}^{C_{b+1}} (1 - R_1)(1 - R_4) \qquad (5)$$

Let

$$\begin{cases} P_1 = F_{A_b}^{B_b} \tau_{b,b+1} \\ P_2 = F_{B_b}^{B_b} \tau_{b,b+1} \tau_{b+1,b} \\ P_3 = E_{A_b,b\sim b+1}^{C_{b+1}} \\ Q = \exp(-k_{b+1} L_{b+1}) \end{cases} \qquad (6)$$

where $k_{b+1}$ is the absorption coefficient of the liquid and $L_{b+1}$ is the thickness of the liquid layer. Establishing a correlation according to formula (4), the following quadratic equation with one unknown is obtained by conversion:

$$(P_3 \cdot R_B \cdot R_C + P_3 \cdot P_2 \cdot R_C)Q^2 + P_1 \cdot Q - P_3 = 0 \qquad (7)$$

For $L_{b+1}$ is the known liquid layer thickness, taking the plus sign $k_{b+1}$ can be derived by

$$k_{b+1} = -\ln(Q)/L_{b+1} \qquad (8)$$

After the reflectance was calculated, the refractive index data of the pure water and quartz [23] and the ionic liquids [24] were obtained. In order to allow for all the possible influential factors during the absorption coefficient inversion, equation (8) was always adopted for calculations.

### 2.2. Propagated Uncertainties in Absorption Coefficient Inversion

In order to analyze the effects of the related influential factors on the absorption coefficient inversion, the square root synthesis method was used to analyze the propagated uncertainties. The basic formula employed is as follows:

$$\frac{\Delta N}{N} = \sqrt{\left(\frac{\partial \ln f}{\partial x_1}\right)^2 \Delta x_1^2 + \left(\frac{\partial \ln f}{\partial x_2}\right)^2 \Delta x_2^2 + \left(\frac{\partial \ln f}{\partial x_3}\right)^2 \Delta x_3^2 + \cdots + \left(\frac{\partial \ln f}{\partial x_n}\right)^2 \Delta x_4^2} \qquad (9)$$

where, $x_1$, $x_2$, $x_3$…$x_n$ are directly measured parameters independent of each other and each of them must be a parameter that is determined through multiple equal-accuracy measurements and contains a random error; $N$ is an indirectly measured parameter ($N=f(x_1, x_2, x_3...x_n)$), and $\Delta N$ is the absolute error of the indirect measurement.

$$\Delta N = \sqrt{\left(\frac{\partial f}{\partial x_1}\right)^2 \Delta x_1^2 + \left(\frac{\partial f}{\partial x_2}\right)^2 \Delta x_2^2 + \left(\frac{\partial f}{\partial x_3}\right)^2 \Delta x_3^2 + \cdots + \left(\frac{\partial f}{\partial x_n}\right)^2 \Delta x_4^2} \quad (10)$$

During the absorption coefficient inversion process, the errors cover that of the transmittance measurement error of the experimental system, the refractive index error of sample tank, the refractive index error of measured liquid and the measurement error of liquid layer thickness.

According to the research findings of Huang [25], the relative errors of the calibrated transmission/reflection spectrum measurement system are not greater than 3%. In the study, the transmittance measurement errors in the two cases including empty sample tank and liquid-containing sample tank were considered to be 3%. The value of the quartz glass sample tank's refractive index was obtained from reference [23]. Considering that there might be a certain difference between the quartz material of the glass tank and that stated in the reference, the relative error of the glass tank's refractive index was assumed to be 1% during the uncertainty propagation analysis. The value of the refractive index of each measured liquid was determined by using the SpectroMaster, a fully automatic high-accuracy refractive index measuring instrument from the Technical Institute of Physics and Chemistry, CAS as well as the minimum deviation angle method, with the relative error smaller than 1.0E-4 (high-accuracy). This was to ensure that, the impact of the liquid's refractive index error was not considered during the analysis. A Vernier caliper with the accuracy of 0.05mm was used to measure the thickness of the liquid layer. The liquid thickness was assumed to be 0.25cm and the liquid layer thickness error was smaller than 2%.

Two measurement cycles were carried out respectively for the two cases of empty sample tank and not-empty sample tank. The measurement errors were represented as $\tau_2$ and $\tau_5$, and the uncertainty propagation formula (11) was used when the absolute error was analyzed during the absorption coefficient measurement:

$$\Delta k_{b+1} = \sqrt{\left(\frac{\partial k_{b+1}}{\partial n_{glass}}\right)^2 \Delta n_{glass}^2 + \left(\frac{\partial k_{b+1}}{\partial \tau_2}\right)^2 \Delta \tau_2^2 + \left(\frac{\partial k_{b+1}}{\partial \tau_5}\right)^2 \Delta \tau_5^2 + \left(\frac{\partial k_{b+1}}{\partial L_{b+1}}\right)^2 \Delta L_{b+1}^2} \quad (11)$$

In case multiple reflections are considered, the following formula can be obtained:

$$k_{b+1} = \ln\left[2P_3(R_B R_C + P_2 R_C)\Big/\left(-P_1 + \sqrt{P_1^2 + 4P_3^2(R_B R_C + P_2 R_C)}\right)\right]\Big/L_{b+1} \quad (12)$$

Substitute $n$ for $n_{glass}$ in the formula (11). The parameter $n$ here still represents the glass's refractive index.

Here if we denoted a parameter $W_P$ as follows

$$W_P = \sqrt{P_1^2 + 4WP_3} \quad (13)$$

The right side of the formula (11) can be obtained based on the following expressions. The absolute values in the terms on the right side of the equation can be expressed as follows

$$\begin{cases} \dfrac{\partial k_{b+1}}{\partial n} = \dfrac{1}{L_{b+1}}\left(\dfrac{1}{W}\dfrac{\partial W}{\partial n} - \dfrac{1}{-P_1+W_P}\left(\dfrac{2\partial P_1}{\partial n}+\dfrac{1}{W_P}\right)\left(P_1\dfrac{\partial P_1}{\partial n}+2P_3\dfrac{\partial W}{\partial n}+2W\dfrac{\partial P_3}{\partial n}\right)\right) \\ \dfrac{\partial k_{b+1}}{\partial \tau_2} = \dfrac{R_B}{L_{b+1}}\left(\dfrac{(W_P-P_1)\partial F_{A_b}^{B_b}/\partial \tau_2 - 2P_3^2 R_B R_C \partial F_{B_b}^{B_b}/\partial \tau_2}{W_P^2 - P_1 W_P} + \dfrac{P_3 R_B R_C \partial F_{B_b}^{B_b}/\partial \tau_2}{W}\right) \\ \dfrac{\partial k_{b+1}}{\partial \tau_5} = \dfrac{1}{L_{b+1}}\left(\dfrac{-4W}{W_P(W_P-P_1)(1-R_1)(1-R_4)}+\dfrac{1}{\tau_5}\right) \\ \dfrac{\partial k_{b+1}}{\partial L_{b+1}} = \dfrac{\ln\left[(W_P-P_1)/2W\right]}{L_{b+1}^2} \end{cases} \quad (14)$$

The absolute error can be obtained by substituting formula (14) into formula (11). Then, the relative error is obtained through dividing the absolute error by the value calculated with formula (12).

## 3. Experimentation
### 3.1. Materials

The materials used in this study were $C_3MImI$ and its four derivatives including $C_4MImI$ (methyl butyl imidazole iodide), $C_5MImI$ (methyl pentyl imidazole iodide), $C_3MImBr$ (methyl propyl imidazole bromide), and $C_3MImBF_4$ (methyl propyl imidazole tetrafluoroborate). All the ionic liquids were purchased from the Lanzhou Institute of Physical Chemistry, Chinese Academy of Sciences.

### 3.2. Principle and Methods

For the incident radiation energy and emission radiation energy, a detector was used with a lock-in amplifier to amplify the measured signal, extract the effective signal and convert it into a corresponding voltage signal. The corresponding value of the voltage was calculated by $U=K\cdot\phi\cdot I_\lambda\cdot\eta_\lambda\cdot\Delta\lambda$. Where $K$ is the number of geometric gathers of the system, $\phi$ is the responsive efficiency of the lock-in amplifier, $I_\lambda$ is the spectral radiation intensity, $\eta_\lambda$ is the spectral responsiveness of the detector, and $\Delta\lambda$ is the bandwidth of the radiation source. When the voltage response of the incident and emitting light source signals was obtained, the transmittance of the sample can be calculated by $T_\lambda=U_i/U_0$. Where $U_i$ is the voltage response of the spectral emitting signal, and $U_0$ is the voltage response of the incident signal. The calibration and error analysis of the transmission spectrum measurement system was carried out according to the scheme previously published by the research group [1] and will not be elaborated here. The literature indicates that the maximum relative error of the calibrated system is no more than 3%.

### 3.3. Apparatus

An existing transmission spectrum measurement and control system was used. In the experiment, its transmittance measurement function was mainly used to measure the light transmittances of the ionic liquids at external voltages as well as their changes. The measurements were done in a dark environment at normal temperature and pressure.

The principle and physical diagram of the transmission spectroscopy used are shown in Fig. 3. The system can be used to measure the transmittance of the liquid layer at different voltages. Secondly, this system can provide a stable and smooth continuous light source, and has the characteristics of good measurement repeatability, a small impact of stray light and high accuracy of signal-to-noise ratio measurement.

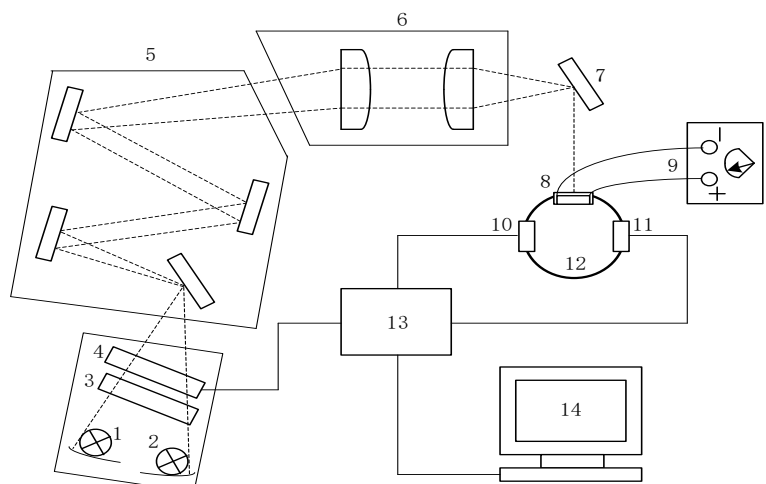

Fig. 3 Illustration of a measurement system for transmission spectrum

Note: 1-Deuterium lamp, 2-Tungsten lamp, 3-Optical filter, 4-Chopper, 5-Monochromator with double grating, 6-Collimating lens, 7-flat mirror, 8-sample holder, 9-Voltage adjustable DC power supply, 10-Silicon detector, 11-InGaAs detector, 12-Integrating sphere, 13-Lock-in amplifier, 14-Data acquisition system.

The experimental apparatus is sketched in Fig. 4. The incident light passes through the measured liquid and the sample tank perpendicularly and is received by the detector. The signal is amplified by the lock-in amplifier, and then the transmittance value is outputted to the computer terminal after the automatic software calculation.

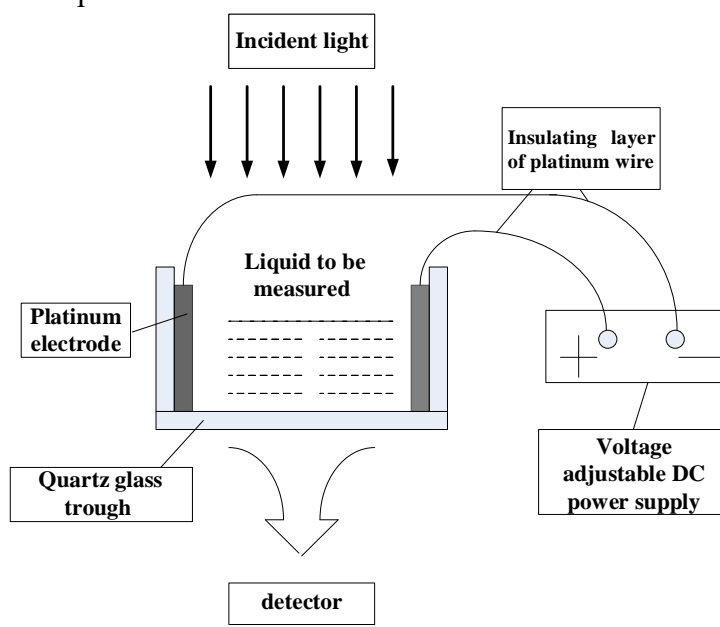

Fig. 4 Electronically controlled ionic liquid transmittance measurement

### 3.4. Procedure

The experimental procedure includes calibration, the transmittance measurement of empty glass tank and ionic liquid. After that, the absorption coefficient inversion and error analysis were conducted. During calibration, the outlet passage of the monochromator was aligned on the same horizontal line as the detector's receiving passage. The grating was adjusted to the appropriate position to ensure that the lock-in amplifier's signal is maximum. During the transmittance measurement of the empty glass, comparison measurements were conducted under the deuterium lamp when no sample was provided and the deuterium lamp in the 380-2500 nm band when a sample was provided. Then, the current response was measured under the tungsten lamp when no sample was provided. Eventually, the transmittance was calculated using the instrument's built-in algorithm.

An appropriate amount of measured liquid using an injector with the 0.025 ml graduation was taken and injected into the sample tank. The spacing between positive and negative electrodes is 5 cm. The liquid was allowed to uniformly spread without any bubbles for 10-20 minutes in the sample tank. During the process, a micrometer was used to make sure that the liquid layer's thickness was 0.25 cm. In case the liquid layer was too thick or too thin, a tiny amount of the liquid was take away from or inject into the sample tank. When the liquid layer became stable, the positive and negative poles of the voltage adjustable DC power supply were connected to the platinum wires of the sample tank respectively. The voltage on the power supply was then set to zero. The sample tank containing the liquid was kept into the transmittance measurement chamber of the spectrograph and the transmittance of the entire apparatus containing the liquid measured. After each set of measurement data was obtained, the power supply voltage was adjusted by slowly rotating the knob to allow the voltage to gradually increase from zero. In order to ensure that the voltage is not high enough to electrolyze the ionic liquid, the experimental measurement voltages are required to include 0 V, 0.5 V, 1 V, 1.5 V, 2 V, 2.5 V, 3 V, and 3.5 V. Given the electrode spacing of 5 cm, the corresponding electric field intensity is 0 V/m, 10 V/m, 20 V/m, 30 V/m, 40 V/m, 50 V/m, 60 V/m and 70 V/m, respectively. The apparatus automatically records each liquid's transmittance versus wavelength curve under each of these voltages. During the voltage increase process, the voltage cannot be decreased at any time. Before each new measurement was made, the system current voltage was kept for three minutes' period. The measurement procedure was done in ascending order of voltage. The experiments were carried out under room temperature conditions (20°C) and one atmosphere (1 atm).

The absorption coefficient inversion process completed in the study was not related to wavelength. It was related to only five factors including overall transmittance, empty sample tank's transmittance, sample tank's refractive index, liquid's refractive index, and liquid layer thickness. In order to analyze the overall error of the system within the space controlled by the five factors, the upper/lower limit analysis method was employed.

The overall transmittance was assumed to change continuously only in the 0.05-0.85 range. The sample tank's spectral transmittance ranges from 0.918 to 0.928 with the lower limit assumed to be 0.918 and the upper limit assumed to be 0.928. Similarly, the quartz sample tank's spectral refractive index ranges from 1.4 to 1.5 whiles the measured liquid's spectral refractive index was assumed to range from 1.05 to 2.0. The liquid layer's thickness was taken 0.25 cm. Based on the above values, the total transmittance is continuously changed, and eight types of the limiting conditions are constructed. The detail of the calculation conditions is listed in Table 1.

The ionic liquid used in this work has a purity of more than 99.5%. Since this paper focuses on the unique phenomenon of the change of the absorption coefficient of ionic liquid under electric field, rather than the calibration of the pure ionic liquid absorption coefficient, the effects of impurities on the spectral curve will not be discussed.

Table 1 Detail of eight typical cases of calculation conditions

|  | Case 1 | Case 2 | Case 3 | Case 4 | Case 5 | Case 6 | Case 7 | Case 8 |
|---|---|---|---|---|---|---|---|---|
| $n_{glass}$ | 1.5 | 1.4 | 1.5 | 1.4 | 1.5 | 1.4 | 1.5 | 1.4 |
| $n_{liquid}$ | 2 | 2 | 1.05 | 1.05 | 2 | 2 | 1.05 | 1.05 |
| $\tau_{glass}$ | 0.928 | 0.928 | 0.928 | 0.928 | 0.918 | 0.918 | 0.918 | 0.918 |

According to the parameters listed in the table, the relationship between the inversion relative error of the absorption coefficient and the corresponding absorption coefficient value with the total transmittance is obtained respectively, as shown in Fig. 5. As can be seen, the relative error of the absorption coefficient measurement increases with the increase of the transmittance. When the transmittance is no less than the lowest sensitivity of the experimental equipment, the larger the absorption coefficient, the smaller the relative measurement error. For the band with larger absorption coefficient, the transmittance can be maintained within the appropriate range by changing the thickness of the liquid layer. In addition, it can be seen that the relative error of absorption coefficient caused by the above factors is less than 15% when the absorption coefficient is greater than 0.5 cm$^{-1}$. With considering the test factors do not change during a set of tests for each liquid to be tested, the above errors are conformity error.

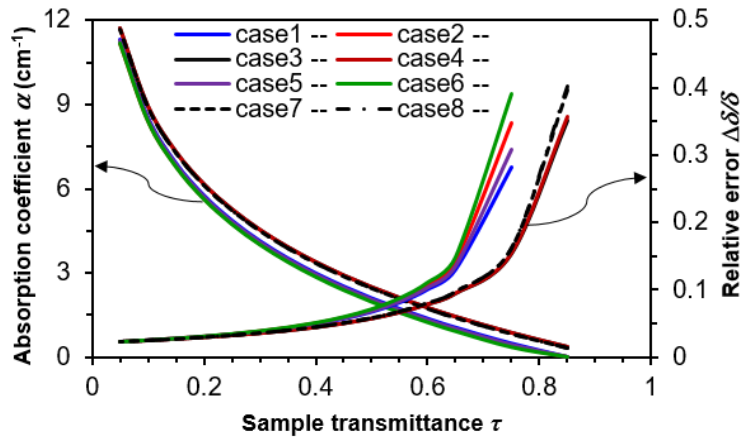

Fig. 5 Illustration of the absorption coefficient and the inversion relative error versus the sample transmittance.

## 4. Results and Discussions

The absorption coefficient values of the five kinds of ionic liquids were calculated according to the measured transmittance values and their absorption coefficient versus wavelength curves at different voltages plotted. Under the electric field, the elementary experiment shows that as long as the voltage does not exceed the threshold of the decomposition of the ionic liquid, the test results are reproducible at the same voltage. However, the voltage thresholds of the decomposition of the ionic liquids before the experiment were unknown. It was ensured that the amount of liquid was constant during the test and the ionic liquid did not break down as the voltage gradually increases, thus the voltage in the test was increased step by step (the step size is 0.5 V).

An illustration of the absorption coefficient of the different ionic liquids at 0 V and 1 V were compared in Fig. 6. As shown in the figure, there are clear differences in the absorption coefficient versus wavelength for the different ionic liquids. The effects of applied voltage on the absorption coefficient are variable for the different ionic liquids also. The changes in absorption coefficients of ionic liquids under atmosphere and under electric field at different wavelength are different. For investigation of these changes in detail, the absorption coefficient maps versus wavelength are provided as follows.

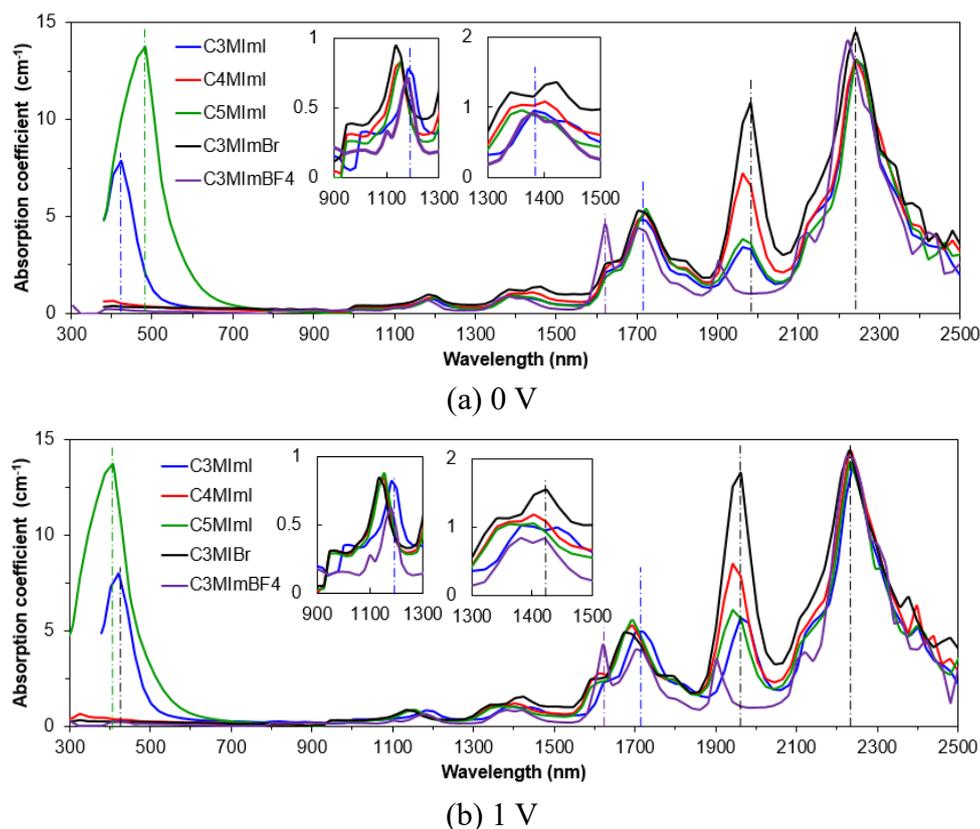

(a) 0 V

(b) 1 V

Fig. 6 Comparison of absorption coefficient of different ionic liquids at 0 V and 1 V

The absorption coefficients of $C_3MImI$ versus wavelength under electric field are shown in Fig. 7. It can be seen that at 1440 nm and 1960 nm, the absorption coefficient increases markedly when the intensity of the applied electric field increases. The absorption coefficient at 1520 nm increases by 114% and that at 1960 nm by 328.1% when the voltage increases from 0 V to 3.5 V. Inversely, the absorption coefficient at two bands of wavelength (400~430 nm, 1675~1735 nm) does not present a clear trend with the applied voltage. Note that the absorption coefficient at 1 V was the largest, whereas that the absorption coefficient at 2 V near to the ultraviolet band was the smallest and at 3.5 V near the infrared band the smallest.

The absorption coefficients of $C_4MImI$ versus wavelength under an applied electric field are shown in Fig. 8. It can be seen that when the wavelength is shorter than 1400nm, the absorption coefficient still does not change considerably with the applied voltage, which is related to electric field intensity. At 1440 nm and 1960 nm, the absorption coefficient increases markedly when the intensity of the applied electric field increases. The absorption coefficient at 1520 nm increases by 99% and that at 1960 nm by 141% when the voltage increases from 0 V to 3.5 V. In addition, it was found that the absorption coefficient changes most dramatically when the voltage increases from 0.5 V to 1 V. However, the absorption coefficient is not dependent upon the applied electric field at two bands of wavelength (400~430 nm, 1700~1740 nm). Note that the absorption coefficient under a 3V voltage is the largest at a band of 400~430 nm, and of that under a 1.5 V voltage is the largest at a band of 1700~1740 nm.

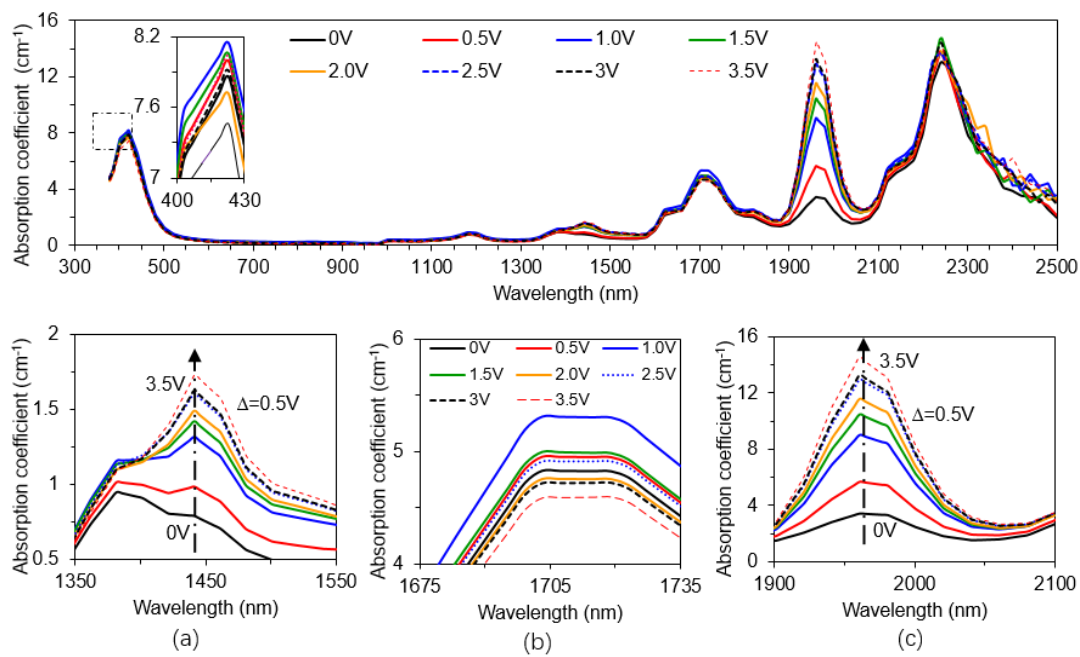

Fig. 7 Absorption coefficient of $C_3MImI$ vs. wavelength curves at different voltages

The absorption coefficients of $C_5MImI$ versus wavelength under an applied electric field are illustrated in Fig. 9. It can be seen that at 1440 nm and 1960 nm, the absorption coefficient increases markedly when the intensity of the applied electric field increases. Furthermore, there were several regions of wavelength for $C_5MImI$, which divided the curves of absorption coefficient versus wavelength at different applied voltages. Note that the absorption coefficient increased clearly with the applied voltage at bands of 1350~1550 nm and 1900~2100 nm. However, this situation no longer exists at bands of 460~490 nm and 1700~1740 nm.

The absorption coefficients of $C_3MImBr$ versus wavelength under an applied electric field are illustrated in Fig. 10. It can be seen that at 1440 nm and 1960 nm, the absorption coefficient increases markedly when the intensity of the applied electric field increases. At 1520 nm, the absorption coefficient values at 0 V, 0.5 V, 1 V, 1.5 V, 2 V, 2.5 V, 3 V, and 3.5 V are respectively 1.018 cm$^{-1}$, 1.092 cm$^{-1}$, 1.315 cm$^{-1}$, 1.504 cm$^{-1}$, 1.643 cm$^{-1}$, 1.775 cm$^{-1}$, 1.953 cm$^{-1}$, and 1.984 cm$^{-1}$. This indicates that the absorption coefficient increased by 91.8% when the voltage increases from 0 V to 3 V. In the above band, the absorption coefficient at 1460 nm is the maximum and increases by 124% when the voltage increases from 0 V to 3 V. At 1960 nm, the absorption coefficient values at 0 V, 0.5 V, 1 V, 1.5 V, 2 V, 2.5 V, 3 V, and 3.5 V are respectively 10.885 cm$^{-1}$,

13.948 cm$^{-1}$, 17.844 cm$^{-1}$, 20.294 cm$^{-1}$, 21.789 cm$^{-1}$, 23.037 cm$^{-1}$, 24.001 cm$^{-1}$, and 24.510 cm$^{-1}$. The absorption coefficient increases by 120% when the voltage increases from 0 V to 3 V.

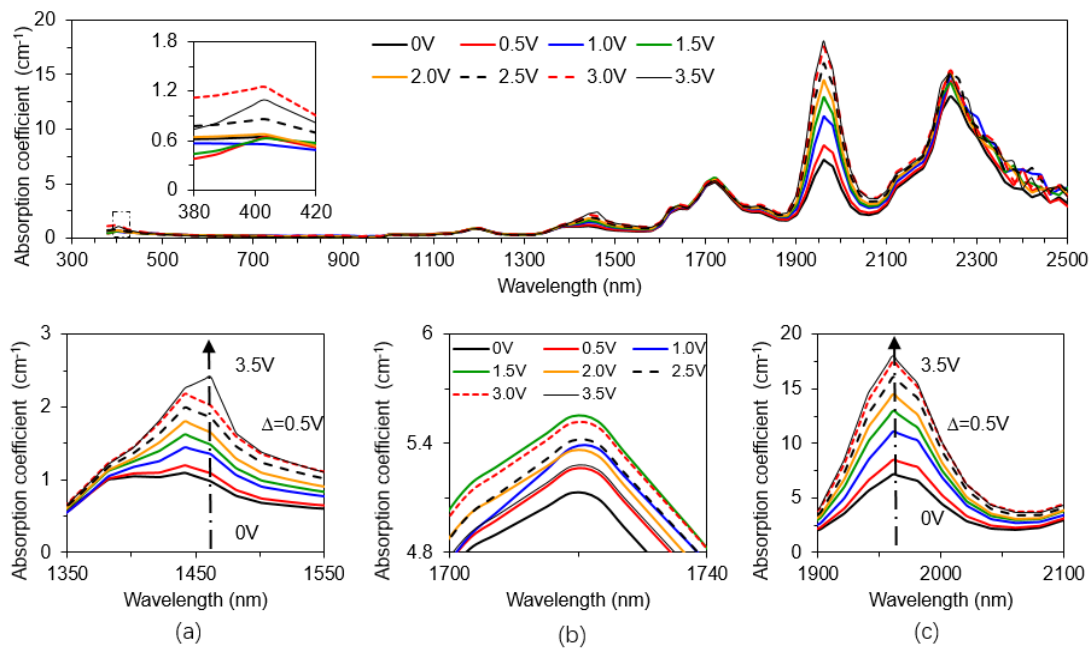

Fig. 8 Absorption coefficient of C$_4$MImI vs. wavelength at different voltages: the upper is an original map, and the lower are local magnification maps.

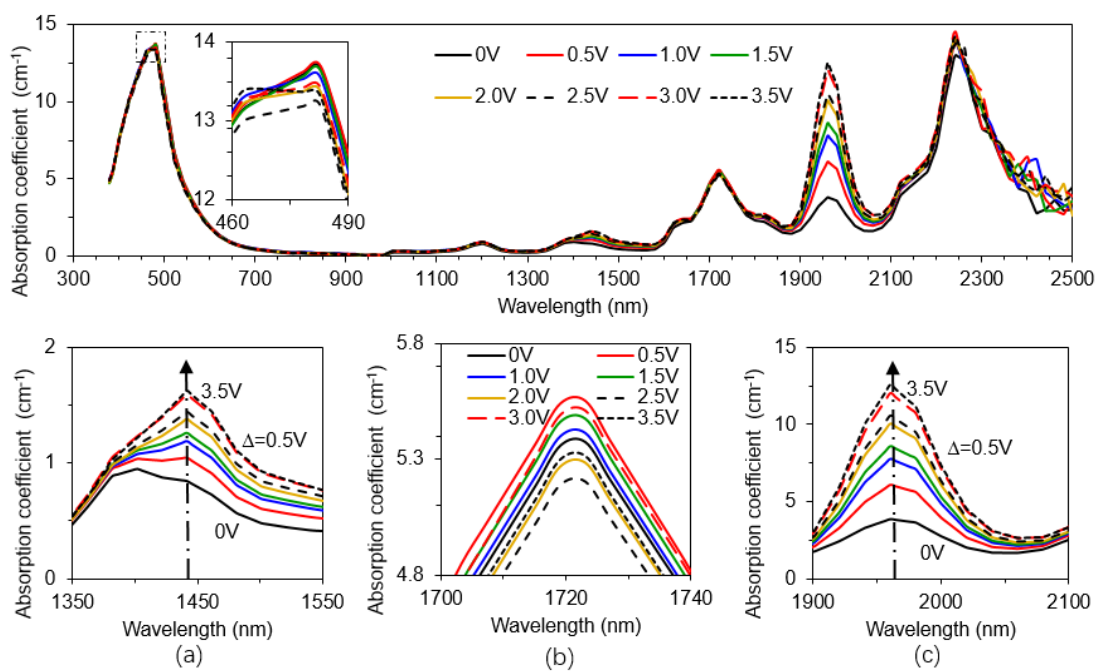

Fig. 9 Absorption coefficient of C$_5$MImI vs. wavelength curves at different voltages

The absorption coefficients of C$_3$MImBF$_4$ versus wavelength under an applied electric field are illustrated in Fig. 11. It can be seen from the figure that the existence of the radical BF$_4$ brought about a spectral structure different from those of the several working media above. In general, the absorption coefficient at 1440 nm does not change considerably and that at 1960nm gradually increases when the voltage increases. Taking 1920 nm for example, the absorption coefficient values at 0 V, 0.5 V, 1 V, 1.5 V, 2 V, and 2.5 V are respectively 2.206 cm$^{-1}$, 2.458 cm$^{-1}$, 3.253 cm$^{-1}$, 3.730 cm$^{-1}$, 4.368 cm$^{-1}$, and 4.736 cm$^{-1}$. It can be seen that during the voltage increase from 0 V to 2.5 V, the absorption coefficient increases by 114%. In addition, it can also be seen that the inversion intended for obtaining absorption coefficient values is impossible due to the sharp reduction in transmittance within the band 1880-1960 nm and beyond 2100 nm when the voltage increases to 3 V or 3.5 V. The samples changed color when the voltage increased to 3 V or 3.5 V. This may be due to high voltage beyond the electrochemical window of the ionic liquid and the high voltage allowed the ionic liquid to be subjected to ionization breakdown. Therefore, the voltage in the experiment had to be controlled within 2.5 V.

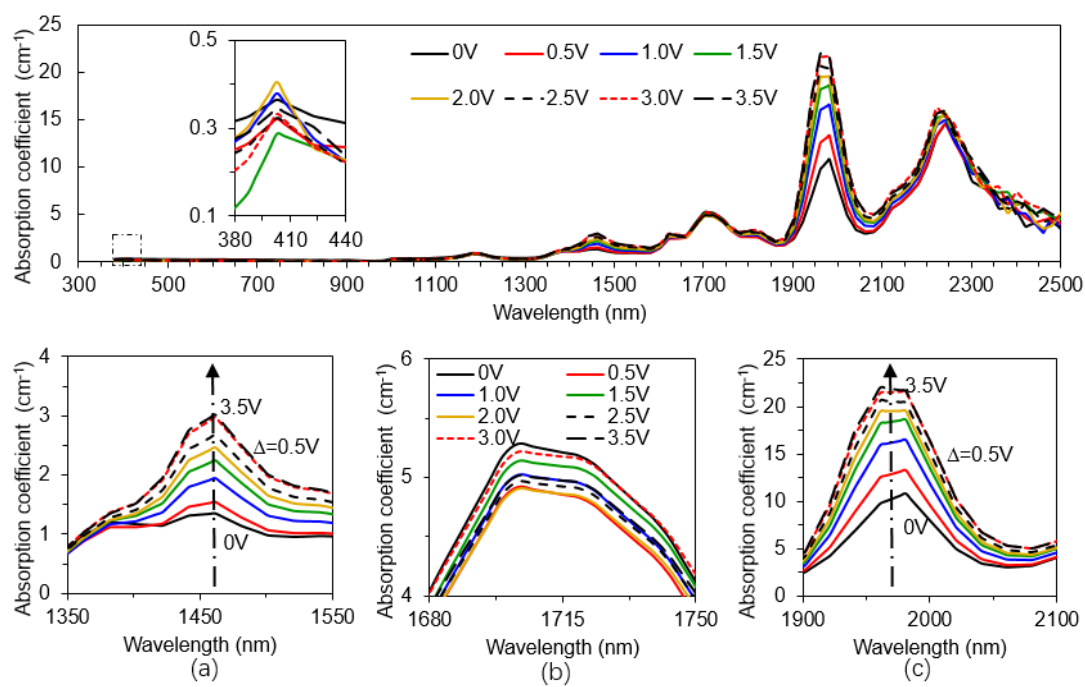

Fig. 10 Absorption coefficient of $C_3MImBr$ vs. wavelength at different voltages

According to a comprehensive analysis of the experimental data, the absorption coefficient rankings by anion substituent when no voltage is applied and the incident light's wavelength of 1520 nm and 1960 nm are obtained as follows: $C_3MImBr > C_3MImI > C_3MImBF_4$. However, the absorption coefficient has different rankings by carbon chain length at different wavelength. It has a following rank $C_4MImI > C_3MImI > C_5MImI$ at 1520 nm, whereas $C_4MImI > C_5MImI > C_3MImI$ at 1960 nm.

In order to compare the absorption coefficient changes of the different ionic liquids under an applied electric field, the ratio of the absorption coefficient $\alpha$ with an applied electric field to the absorption coefficient $\alpha_0$ without an applied electric field defined as the effect coefficient of the applied electric field were determined. The ratio representing the absorption coefficient at given electric field and that of 0 V/m reflects the degree of impact of the electric field on the absorption coefficient. Thus, the changes of the absorption coefficient values at 1520 nm and 1960 nm under different uniform electric field were plotted in Fig. 12. The ionic liquid absorption coefficient versus applied voltage was analyzed. The absorption coefficient changed multiples times at the same voltage and are ranked by the carbon chain length as follows: $C_3MImI > C_5MImI > C_4MImI$. Besides, the absorption coefficients change multiples times at the same voltage ranked by the anion type as follows: $C_3MImI > C_3MImBr > C_3MImBF_4$.

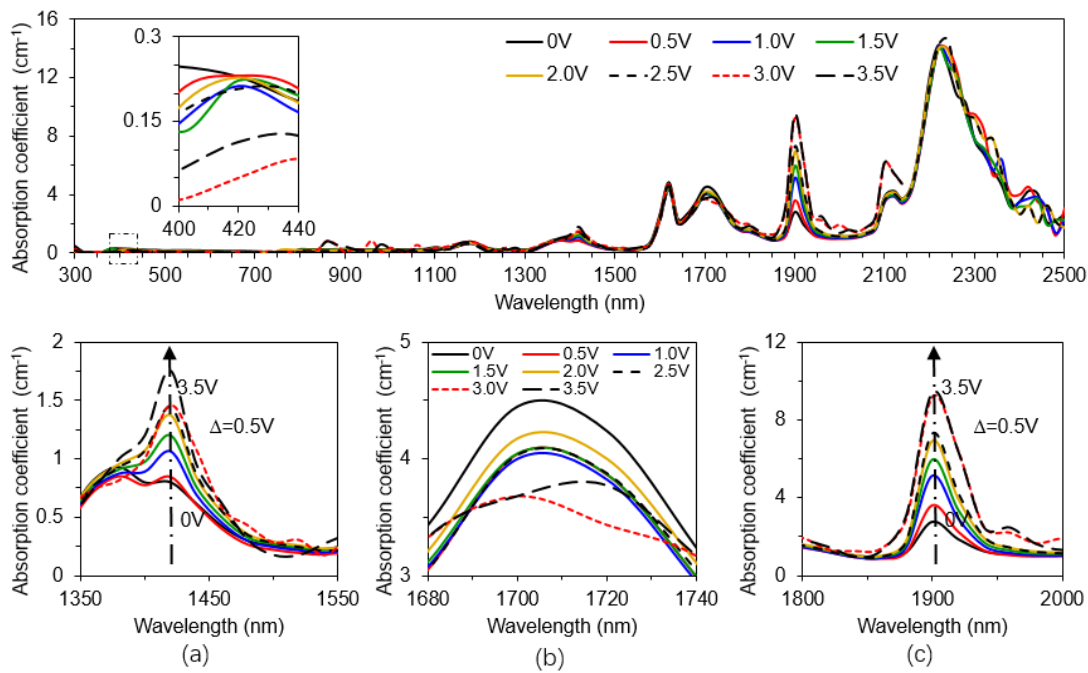

Fig. 11 Absorption coefficient of $C_3MImBF_4$ vs. wavelength at different voltages

  The above curves show that the absorption coefficient change with the voltage is largely dramatic (steep curve) when the electric field is lower than 20V/m and tends to be gradually gentler (gentle curve) when the voltage is higher than 20V/m. This is mainly because the ion motions (ion chain formation, ion clustering, etc.), the anion and cation concentrations are changed or affected initially at low electric field. Importantly, the distribution pattern and motion state of the ions tend to be steady at high electric field (before the ionic liquid is not decomposed). Besides, many alien charges both at electrodes lead to higher ion migration resistances toward both electrodes, thus inhibiting further arrangement and migration of the ions as well as reducing the change of the absorption coefficient with the intensity of the applied electric field.

  Differences of absorption coefficients versus voltage for several ionic liquids may be accounted for the physicochemical properties. The bond energy between carbon chain and imidazole ring in $C_3MimI$ is lower than those in $C_5MImI$ and $C_4MImI$. Under an electric field, $C_3MimI$ having the smaller ion migration resistances is affected significantly by the applied electric field. When no electric field is applied, the $C_4MImI$'s absorption coefficient is higher than those of $C_3MImI$ and $C_5MImI$. After that, the effect of applied electric field on $C_4MImI$ is not significant. Under an electric field, the separation of the cation-imidazole ring of $C_3MImI$ is easier because the cations are difficult to oxidize, and therefore more significantly affected by the electric field. In $C_3MImBF_4$ whose cations are easy to oxidize, the separation of the cation-imidazole ring is relatively hard under an electric field, and therefore the impact of the electric field on $C_3MImBF_4$ is smaller.

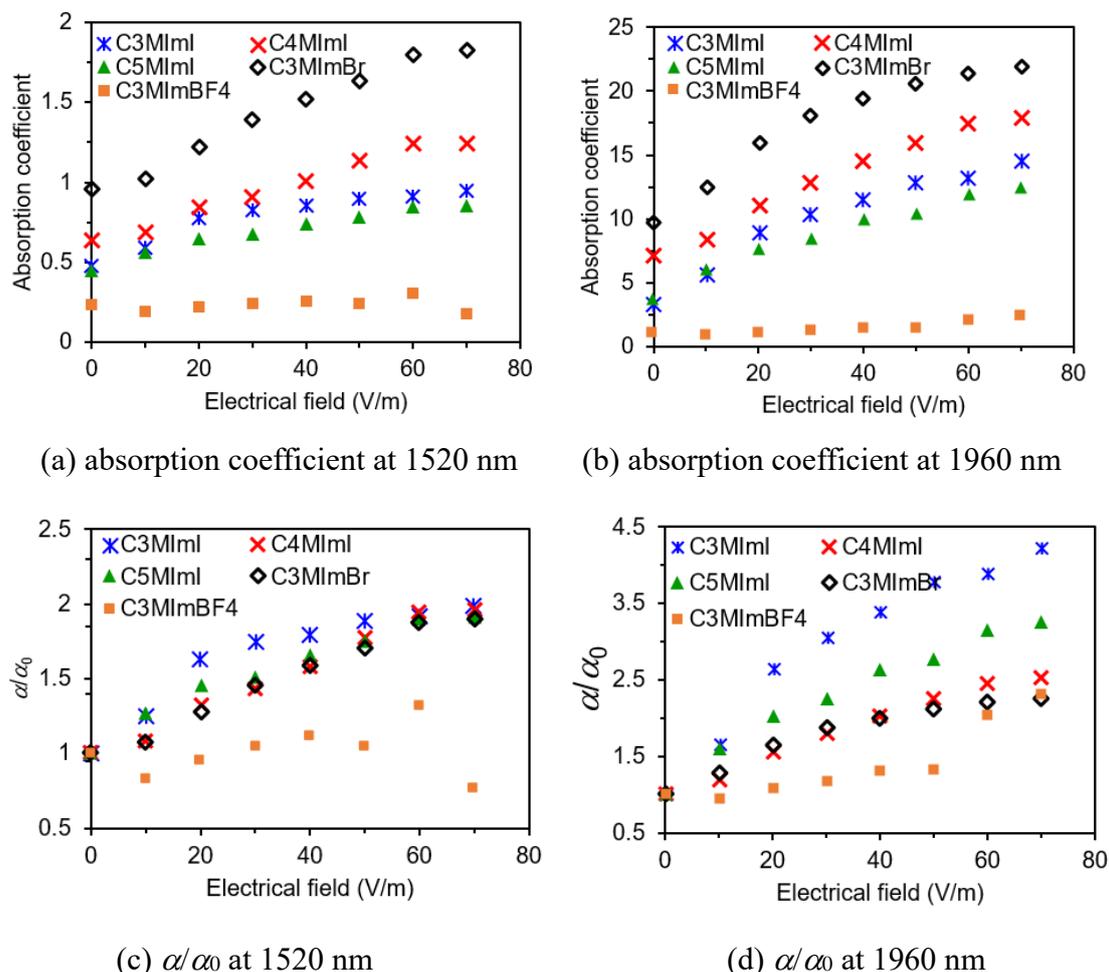

(a) absorption coefficient at 1520 nm     (b) absorption coefficient at 1960 nm

(c) $\alpha/\alpha_0$ at 1520 nm     (d) $\alpha/\alpha_0$ at 1960 nm

Fig. 12 Comparison of absorption coefficient and the electric effects for two wavelengths at different voltages

## 5. Conclusions

The present study investigated the light transmission properties of ionic liquids in a uniform electric field. A rectangular glass tank with electrodes on both sides was designed to measure the transmittance. An experimental program for measuring light absorption properties of ionic liquids within ultraviolet, visible, and near-infrared spectra was also developed. Five kinds of imidazole ionic liquids' spectral absorption coefficient versus electric field intensity laws were identified and illustrated. Multiple refractions and reflections were considered during the process of absorption coefficient inversion. An analysis of propagated uncertainties involved in the measurement results carried out by using the uncertainty propagation formula for absorption coefficient and determining the error range of the inversion results. The main conclusions obtained are as follows:

(1) Near 1520 nm and 1920 nm, the absorption coefficient gradually increases when the voltage increases. Specifically, the peak value of absorption coefficient occurs at 1920nm. In the short-wavelength range, the absorption coefficient changes very slightly with the voltage range. Within the above bands, the descending-order rankings in absorption coefficient's susceptibility to the applied electric field by carbon chain length and cation are respectively as follows: $C_3MImI>C_5MImI>C_4MImI$ and $C_3MImI>C_3MImBr>C_3MImBF_4$.

(2) The accuracy of the absorption coefficients measurement was affected by the system's inherent transmittance measurement error, the sample tank's refractive index error, the liquid layer's refractive index error, and the liquid layer thickness error. When the relative error of the transmittance measurement was smaller than 2.4%, the relative errors of the glass tank's refractive index and the liquid's refractive index were smaller than 1% and 0.1%, respectively. The liquid thickness-induced error was also smaller than 2%, whiles the overall transmittance was lower than 65% and higher than 2%. However, the absorption coefficient relative error caused by the above factors was controlled within 20%.

(3) The absorption coefficient changes dramatically with the voltage in the beginning and then tends to change slightly when the voltage continues to increase. This proves that there is a specific connection between ion migration and absorption coefficient change.

The study can provide experimental means and data support for further mechanism analyses and theoretical simulations of ionic liquids' electro-optical effect. It provides strong support for the electro-optical regulation technologies applied in a number of fields such as optical communication, optical sensing, optical displaying, high-power solid laser, smart glass, solar PV generation, etc. It also lays the foundations for further exploring the electro-optical regulation mechanisms and capabilities of ionic liquids, ionic liquid-like soft materials, and metamaterials.

**References**
1. Hu W, Zhang L, Cao H, et al. *2010 Phys. Chem. Chem. Phys.* **11** 2632
2. Wang F H, E.Itkis M E, Bekyarova E, et al. *2013 Nat. Photonics* **7** 459
3. He B, Zhang C H, Ding A *2017 Chin. Phys. B* **26** 126102
4. Liu H X, Man R L, Zheng B S, Wang Z X and Yi P G *2014 Chinese J Chem. Phys.* **27** 144
5. Lu L Y, Zhang Y J, Chen J J, Tong Z H. *2017 Chinese J Chem. Phys.* **30** 423
6. Ding S, Wei L G, Li K L, Ma Y C. *2016 Chinese J Chem. Phys.* **29** 497
7. Du L K, Geng C H, Zhang D J, Lan Z G, and Liu C B. *2016 J Phys. Chem. B* **120** 6721
8. Song Y Y, Tian P, and Yu L *2013 Adv. Mater. Res.* **772** 219
9. Yang Q. *2014 Appl. Mech. Mater.* **685** 106
10. Yu L, Tian P. *2014 Appl. Mech. Mater.* **685** 98
11. Paul A, Mandal P K, Samanta A *2005 J Phys. Chem. B* **18** 9148
12. Paul A, Mandal P K, Samanta A *2005 Chem. Phys. Lett.* **4-6** 375
13. Paul A, Mandal P K, Samanta A *2006 J Chem. Sci.* **4** 335
14. Jo T S, Koh J J, Han H, et al. *2013 Mater. Chem. Phys.* **2-3** 901
15. Zhang L, Liu J, He G D, Ye Z C, Fang X M, Zhang Z G. *2014 Sol. Energy Mater. Sol. Cells* **130** 521

16. Ansari S A, Liu L S, Rao L. *2015 Dalton Trans.* **6** 2907
17. Hu W, Zhao H, Song L, Yang Z, Cao H, Cheng Z, Liu Q, Yang H. *2010 Adv. Mater.* **4** 468
18. Nakano M, Shibuya K, Ogawa N, Hatano T *2013 Appl. Phys. Lett.* 15 153503
19. Reddy B N, Deepa M. *2013 Polymer* 21 5801
20. He X D, Shao Q F, Cao P F, Kong W J, Sun J Q, Zhang X P, Deng Y Q. *2015 Lab Chip* **5** 1311
21. Fu T R, Liu J F *2015 Infrared Phys. Techn.* **69** 88
22. Lou J F *2002 Transient coupled heat transfer in multilayer absorbing-scattering composite with specular and diffuse reflection properties (Ph.D. Dissertation) (Harbin: Harbin industrial University) (in Chinese)*
23. Lide D R *1989 CRC Handbook of Chemistry and Physics, 69th edn. (Boca Raton: CRC Press) pp: E382*
24. Zhou J *2015 Research on the optical transmission characteristics and electrical modulation mechanism of electro-optic material under multi field effect (Ph.D. Dissertation) (Harbin: Harbin industrial University) (in Chinese)*
25. Huang X *2016 Photo-thermal characteristics analysis for concentrating radiation thermal decomposition process based on ferrite (Ph.D. Dissertation) (Harbin: Harbin industrial University) (in Chinese)*